\documentclass[twocolumn,twocolappendix]{aastex631}
\usepackage{amsmath}
\usepackage{amsfonts}
\usepackage{graphicx}
\usepackage{rotating}
\usepackage{natbib}
\usepackage{txfonts}
\usepackage{color}
\usepackage{wrapfig}
\usepackage{amssymb}
\usepackage{color}
\usepackage{colortbl}
\usepackage{mhchem}

\shorttitle{Peculiar Icy Objects with ALMA} 
\shortauthors{T. Shimonishi et al.} 

\begin{document}

\title{ALMA Observations of Peculiar Embedded Icy Objects} 

\correspondingauthor{Takashi Shimonishi} 
\email{shimonishi@env.sc.niigata-u.ac.jp} 

\author[0000-0002-0095-3624]{Takashi Shimonishi} 
\affiliation{Institute of Science and Technology, Niigata University, Ikarashi-ninocho 8050, Nishi-ku, Niigata 950-2181, Japan}

\author[0000-0002-8234-6747]{Takashi Onaka} 
\affiliation{Department of Astronomy, Graduate School of Science, The University of Tokyo, 7-3-1 Hongo, Bunkyo-ku, Tokyo 113-0033, Japan}

\author[0000-0001-7641-5497]{Itsuki Sakon} 
\affiliation{Institute of Astronomy, the University of Tokyo, 2-21-1 Osawa, Mitaka, Tokyo 181-0015, Japan}

\begin{abstract}
We report the results of molecular line observations with the Atacama Large Millimeter/submillimeter Array (ALMA) towards two peculiar icy objects, which were discovered serendipitously by infrared spectroscopic survey of the Galactic plane with the AKARI satellite. 
Previous infrared observations have reported that both objects show deep ice and dust absorption features that are often seen in embedded young stellar objects (YSOs) or background stars sitting behind dense clouds, however, they are located neither in known star-forming regions nor in known dense clouds. 
Their infrared spectral energy distributions (SEDs) show a peak around 5 $\mu$m, which are incompatible with existing SED models of typical embedded YSOs. 
The present ALMA observations have detected compact emission of CO(3-2) and SiO(8-7) at the positions of the icy objects. 
The observed large column ratios of gas-phase SiO/CO ($\sim$10$^{-3}$) in both objects, as well as their broad line widths (8-14 km s$^{-1}$), imply that they are associated with shocked gas. 
Although a large dust extinction ($A_V$ $\sim$100 mag) is expected from their deep dust/ice absorption, no dust continuum emission is detected, which would suggest a large beam dilution effect due to their compact source sizes. 
Their systemic velocities are clearly separated from the surrounding CO clouds, suggesting that they are isolated. 
The characteristics of their SEDs, the presence of deep ice/dust absorption features, compact source size, and SiO-dominated broad molecular line emission, cannot easily be accounted for by any of known interstellar ice-absorption sources. 
They may represent a previously unknown type of isolated icy objects. 
\end{abstract} 


\section{Introduction} \label{sec_intro} 
Various kinds of ice species are formed in dense and cold environments in the interstellar medium (ISM). 
They play a significant role in the chemical evolution of star-forming regions and formation of planetary systems \citep[e.g.,][and references therein]{Boo15}.  
Their absorption features, which are observed in the infrared regime, provide useful information on the interstellar chemistry and the nature of the objects associated with ice absorption. 

Recently, the discovery of two peculiar objects with strong ice absorption is reported based on a multi-object infrared spectroscopic survey of the Galactic plane with the Infrared Camera \citep[IRC;][]{TON07} on board the AKARI satellite \citep{Ona21}. 
However, to date, the nature of these objects are still an enigma. 
Both objects (hereafter Object 1 and 2) show deep ice absorption features, but are located neither in known star-forming regions nor in known dense clouds. 
Deep ice absorption features are typical characteristics of embedded young stellar objects (YSOs) \citep[e.g.,][]{Tie84, Dar99, Gib04, Boo04, Pon08, Obe11, ST10, Ona22, Yan22, Roc24}. 
However, both objects have very blue color in 8--24 $\mu$m; their spectral energy distributions (SEDs) peak at around 4--5 $\mu$m and decrease toward longer wavelengths \citep[see Fig.3 in][]{Ona21}. 
Such infrared SEDs are incompatible with standard SED models of embedded YSOs, and could be better explained by background stars sitting behind a dense could with large extinction, where deep ice absorption features are also commonly detected \citep[e.g.,][]{Whi88, Whi07, Kne05, Boo11, Nob13, Nob17, McC23}. 
However, the observed reddening is localized at the position of the icy objects and no extended dark clouds are listed at their positions in available dark cloud catalogs \citep{Per09, Dob05}, which is consistent with optical detection of adjacent stars. 
The absence of known dense clouds towards the two objects implies the presence of very compact, dense clouds of unknown type, if they are background stars. 

Besides embedded YSOs, ice absorption features are also detected in edge-on/face-on protoplanetary disks \citep[e.g.,][]{Pon05, Ter07, Hon09, Hon16, Aik12, Str23}, mass-losing Asymptotic Giant Branch (AGB) stars \citep[i.e. OH/IR stars, e.g.,][]{Syl99, Jus06, Suh13}, central regions of nearby starburst galaxies \citep[e.g.,][]{Spo02, Spo03, Yam13, Yam15}, ultraluminous infrared galaxies (ULIRGs) \citep[e.g.,][]{Ima06, Ima08, Ris06}, and high-redshift galaxies \citep[e.g.,][]{Saj09}. 
However, the infrared characteristics of the two peculiar icy objects do not match those of any known ice-absorption sources. 

High-spatial-resolution submillimeter observations are a powerful tool to diagnose their nature. 
Previous AKARI observations suggest that the warm (60 K) amorphous H$_2$O ice better fits to the 3 $\mu$m ice profile of Object 1 \citep{Ona21}. 
Tentative detection of high-temperature CO gas ($\sim$150 K) and crystalline silicate feature is also reported. 
These characteristics may be signs of thermal processing of dust and ices (e.g., heating by the central protostar or protostellar outflows/jets). 
The spectral resolution of the AKARI/IRC is, however, not sufficient to draw a definite conclusion on the radiation environment of the objects.  
If the objects are associated with the protostar and already heated up its envelope, then they will harbor high-temperature and dense gas around the protostar (i.e. hot molecular core). 
In this case, submillimeter observations will detect high-excitation transitions of hot-core tracers such as CH$_3$OH and SO$_2$. 
If they are background stars, there must be dense and very compact clouds in the line-of-sight of each object, which have eluded past dense cloud surveys in our Galaxy, or ice species may grow in unknown processes in tenuous clouds. 
Identification of their true nature will make a significant impact on our understanding of the ice chemistry in the ISM. 

In this paper, we report the results of molecular line observations with the Atacama Large Millimeter/submillimeter Array (ALMA) towards these peculiar icy objects for better characterization and understanding of the objects. 
The observations and data reduction are described in Section \ref{sec_tarobsred}, and the results are presented in Section \ref{sec_res}. 
A possible nature of the objects is discussed in Section \ref{sec_disc} and a summary is given in Section \ref{sec_sum}. 


\begin{figure}[tp!]
\begin{center}
\includegraphics[width=8.3cm]{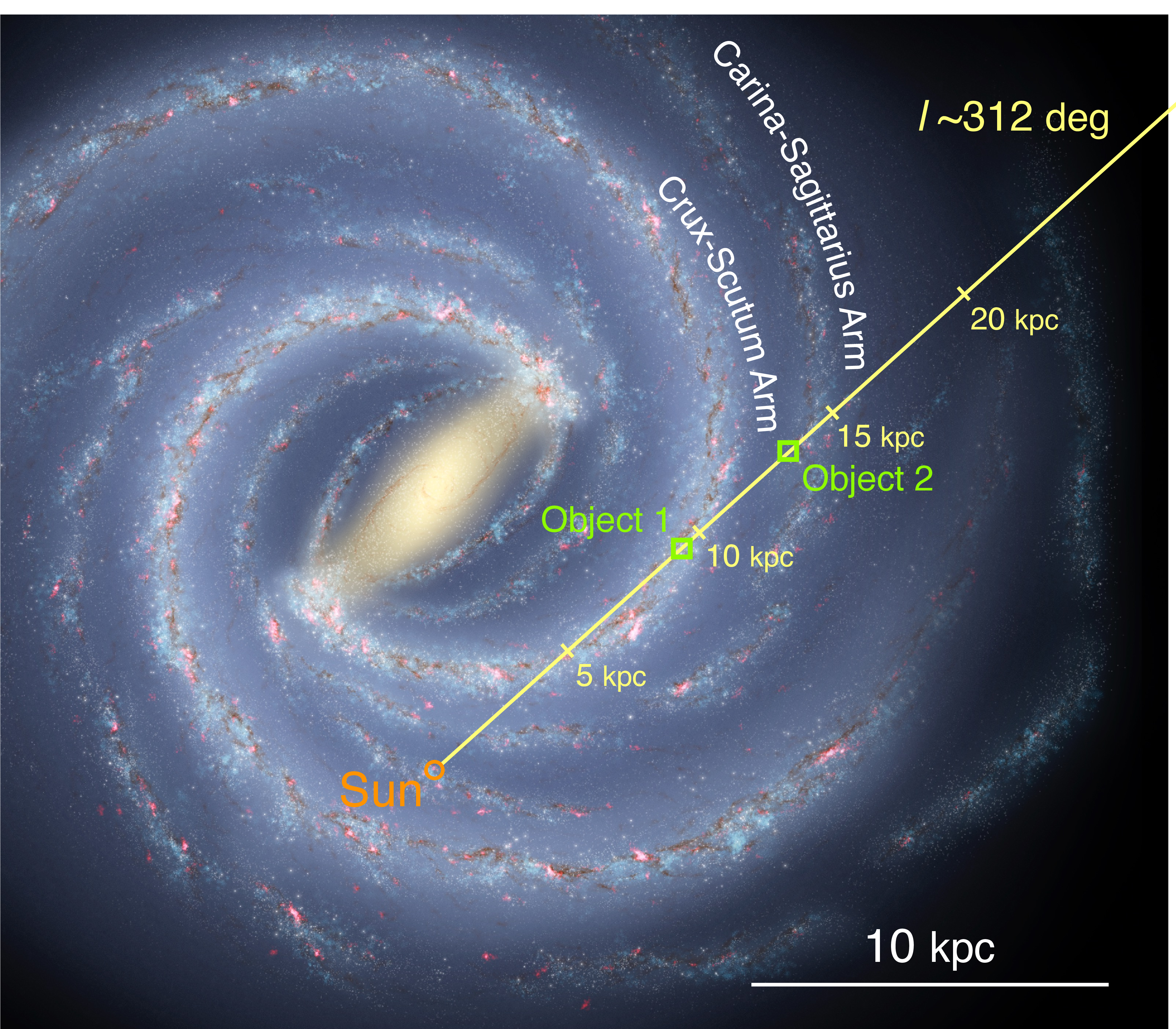}
\caption{
Direction of Object 1 and 2 (solid yellow line). 
The green squares represent their possible locations based on the kinematic distance estimates (see Section \ref{sec_disc3} and \ref{sec_disc4}). 
The background is an artist's conception of the Milky Way (R. Hurt/NASA/JPL-Caltech/ESO). 
}
\label{direction}
\end{center}
\end{figure}

\section{Observations and data reduction} \label{sec_tarobsred} 
\subsection{Targets and Observations} \label{sec_obs} 
The target objects are located in the direction toward the Crux-Scutum arm (Figure \ref{direction}). 
Positions of the infrared point-sources measured by AKARI are RA = 14$^\mathrm{h}$4$^\mathrm{m}$13$\fs$2 and Dec = -61$\arcdeg$12$\arcmin$40$\farcs$1 for Object 1 and RA = 14$^\mathrm{h}$4$^\mathrm{m}$20$\fs$2 and Dec = -61$\arcdeg$15$\arcmin$49$\farcs$5 for Object 2 (IRCS), respectively. 
Their galactic longitude and latitude are ($l$, $b$) = (311.58, 0.43) for Object 1 and (311.58, 0.38) for Object 2, respectively. 
The telescopes were pointed to these infrared positions. 

Observations were carried out with ALMA in April 2022 as a Cycle 8 program (2021.1.01126.S, PI: T. Shimonishi). 
Both the 12 m main array and 7 m Morita array (Atacama Compact Array, ACA) were used. 
The total number of 12 m antennas was 45, where the minimum--maximum baseline lengths were 14.6--313.7 m, while 10 antennas were used for the ACA observations with the minimum--maximum baseline lengths of 8.9--45.0 m. 
The on-source integration time per each target is 5.5 minutes with the 12 m array and 39.3 minutes with the ACA. 
The sky frequency of 344.3--345.2, 345.2--346.1, 346.1--348.0, 356.4--358.3, 358.2--360.1 GHz, are covered by five spectral windows with a velocity resolution of 0.81--0.85 km s$^{-1}$. 
The same spectral setup was used for the 12 m array and ACA observations. 

\begin{deluxetable*}{ l c c c c c c c }
\tablecaption{Line Parameters of Object 1 and 2 \label{tab_lines}}
\tabletypesize{\small} 
\tablehead{
\colhead{Molecule}   & \colhead{Transition}   &  \colhead{$E_{u}$} & \colhead{Frequency} &      \colhead{$T_{br}$}   &   \colhead{$\Delta$$V$}  &   \colhead{$\int T_{br} dV$}  & \colhead{$V_{LSR}$}    \\
\colhead{ }          & \colhead{ }            &  \colhead{(K)}     & \colhead{(GHz)}     &       \colhead{(K)}       &   \colhead{(km/s)}       &   \colhead{(K km/s)}          &  \colhead{(km/s)}            
}
\startdata
Object 1      &           &            &                &                  &                   &                  &            \\
 CO        &  3--2  &   33   & 345.79599    &  0.39 $\pm$  0.04   &   9.9 $\pm$ 1.4   & 3.8 $\pm$ 0.4   & -26.0  \\
 SiO       &  8--7  &   75   & 347.33058    &  0.17 $\pm$  0.04   &   9.9 $\pm$ 2.7   & 1.7 $\pm$ 0.4   & -24.9  \\
\hline
Object 2     &            &            &                &                  &                  &                    &               \\
 CO       &  3--2  &   33    & 345.79599    &  0.18 $\pm$  0.03   &   14.3 $\pm$ 3.1  & 2.7 $\pm$ 0.4    & 29.3   \\
 SiO      &  8--7  &   75    & 347.33058    &  0.10 $\pm$  0.04   &   8.3 $\pm$ 3.7   & 0.9 $\pm$ 0.3    &  30.9  \\
\enddata
\end{deluxetable*}

\subsection{Data reduction} \label{sec_red} 
\begin{figure*}[tp!]
\begin{center}
\includegraphics[width=18.0cm]{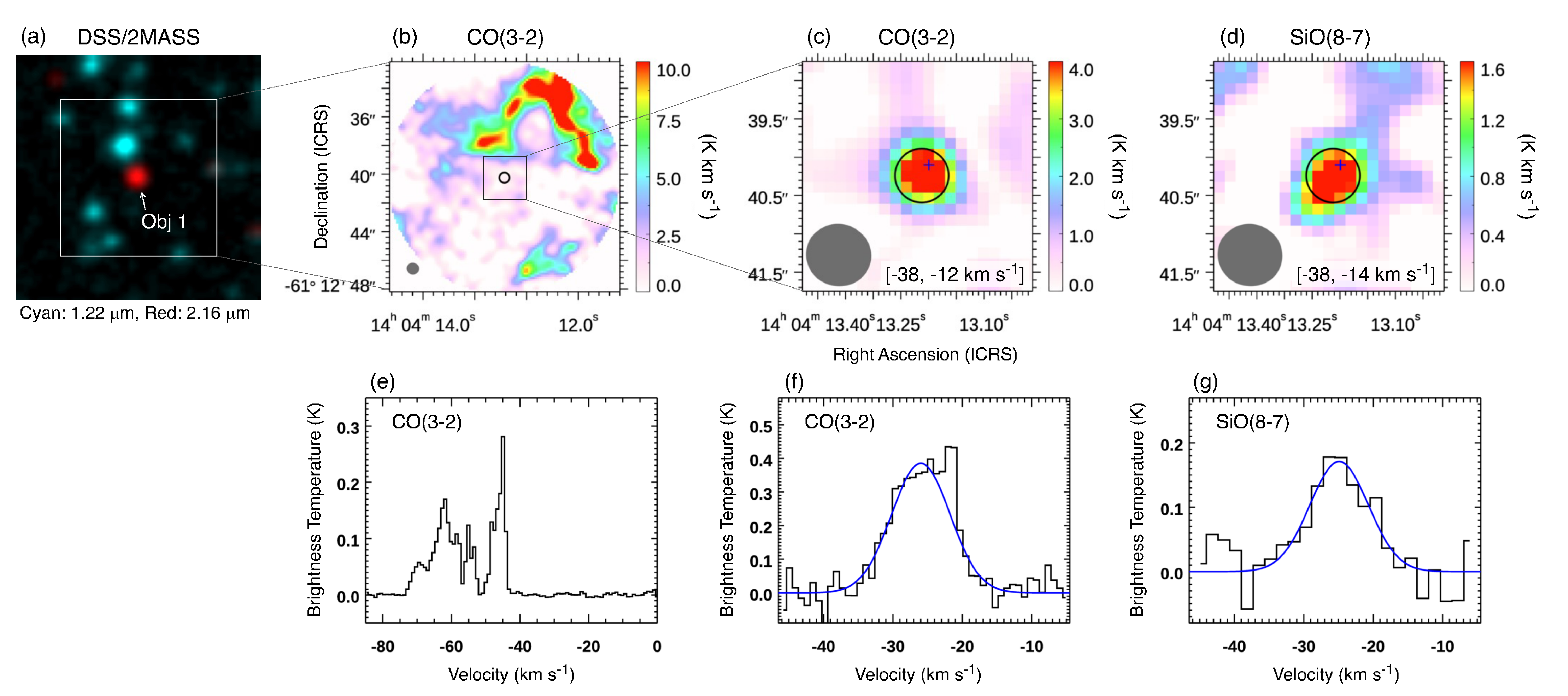}
\caption{
ALMA observations of Object 1. 
(a) Composite two-color image (cyan: 0.61 $\mu$m, red: 2.16 $\mu$m) based on the DSS and 2MASS data. 
The white square indicates the field-of-view of the ALMA observations. 
(b) Integrated intensity distribution of CO(3--2) in a whole field-of-view (16$\arcsec$). 
The image is constructed by integrating the spectral data in a velocity range from $-$75 to $-$40 km s$^{-1}$. 
The panel (e) shows the CO spectrum extracted from the whole area. 
(c, d) The zoom-up images of Object 1 in CO(3--2) and SiO(8--7). 
The images are constructed by integrating spectral data in the velocity range where the emission is detected (CO: $-$38 to $-$12 km s$^{-1}$, SiO: $-$38 to $-$14 km s$^{-1}$). 
The black circle indicates the region of the spectral extraction. 
The blue cross indicates the position of the infrared source, where rich ice/dust absorption bands are detected. 
The gray ellipse represents the synthesized beam size ($\sim$0$\farcs$8 $\times$ 0$\farcs$9). 
The panels (f) and (g) show the CO and SiO spectra extracted from the 0$\farcs$7 diameter region centered at the emission peak. 
The blue curve represents the result of the Gaussian fit. 
}
\label{obj1}
\end{center}
\end{figure*}

\begin{figure*}[tp!]
\begin{center}
\includegraphics[width=18.0cm]{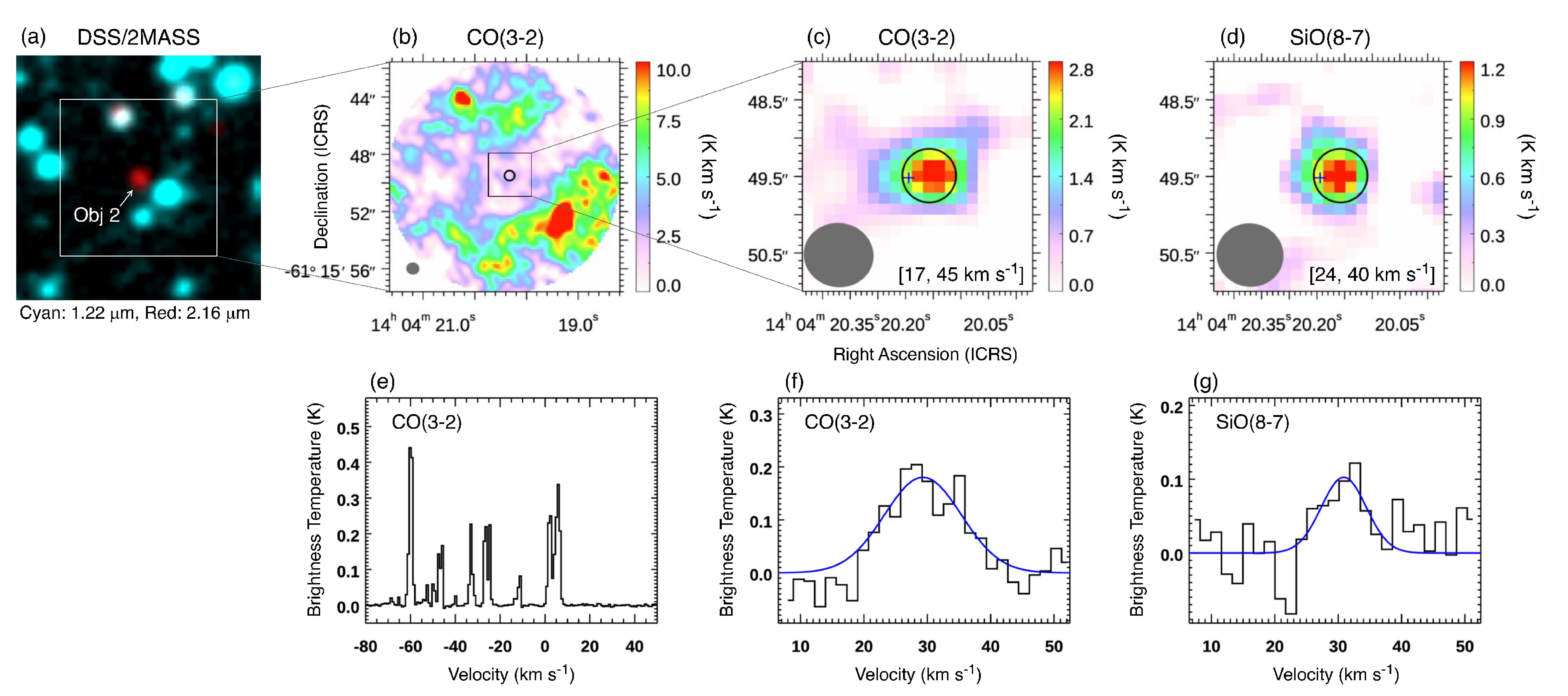}
\caption{
The same as in Figure \ref{obj1}, but for Object 2. 
The integrated intensity distribution of CO(3--2) in the panel (b) was constructed by integrating the spectral data in a velocity range from $-$70 km $s^{-1}$ to $+$10 km $s^{-1}$. 
For those in panels (c) and (d), the images were constructed by integrating spectral data in the velocity range where the emission is detected (CO: 17 to 45 km s$^{-1}$, SiO: 24 to 40 km s$^{-1}$). 
}
\label{obj2}
\end{center}
\end{figure*}

Raw data is processed with the \textit{Common Astronomy Software Applications} (CASA) package \citep[version 6.2.1.7,][]{CASA2022}. 
The CASA task $\mathtt{tclean}$ is used for imaging, and the masking is done using the auto-multithresh algorithm described in \citet{Kep20}. 
The 12 m array data were combined with the ACA data within the $\mathtt{tclean}$ task. 
The continuum emission is subtracted from the spectral data using the $\mathtt{uvcontsub}$ task in CASA. 
The synthesized beam size at the frequency of the CO(3-2) transition is 0$\farcs$85--0$\farcs$91 $\times$ 0$\farcs$81--0$\farcs$84 with the Briggs weighting and a robustness parameter of 0.5. 
We also tried $\mathtt{tclean}$ with a robustness parameter of 2.0 (natural weighting), but the signal-to-noise ratio of emission lines in the target sources did not change significantly. 
The maximum recoverable scale is about 19$\arcsec$. 
The synthesized image is corrected for the primary beam pattern using the $\mathtt{impbcor}$ task in CASA. 
The aggregate continuum image is constructed by selecting line-free channels. 
The field-of-view of the ALMA observations is shown in the panel (a) of Figures \ref{obj1}--\ref{obj2}.

\section{Results and analysis} \label{sec_res} 
We detect compact emission of CO(3--2, $E_{u}$\footnote{Upper state energy.} = 33 K) and SiO(8--7, $E_{u}$ = 75 K) at the positions of Object 1 and 2 (Figures \ref{obj1}--\ref{obj2} c, d, f, g). 
Emission peaks of CO and SiO are located at RA = 14$^\mathrm{h}$4$^\mathrm{m}$13$\fs$213 and Dec = -61$\arcdeg$12$\arcmin$40$\farcs$24 for Object 1 and RA = 14$^\mathrm{h}$4$^\mathrm{m}$20$\fs$158 and Dec = -61$\arcdeg$15$\arcmin$49$\farcs$49 for Object 2 (ICRS), which agree with the positions of the infrared sources. 

The CO and SiO emission are not spatially resolved with the present angular resolution. 
Their spectra are extracted from a 0$\farcs$7 diameter circular area centered at the emission peak. 
The rms noise level is 0.04--0.05 K in the frequency regions of the CO and SiO emission lines. 
We estimate the peak brightness temperature ($T_{br}$), the full-width at half-maximum (FWHM, $\Delta$$V$), the LSR velocity ($V_{LSR}$), and the integrated intensity ($\int T_{br} dV$) for each line with the Gaussian fitting, which are summarized in Table \ref{tab_lines}. 
The systemic velocities of the molecular gas associated with Object 1 and 2 are clearly different from each other ($V_{LSR}$ $\sim$$-$25 km s$^{-1}$ for Object 1 and $V_{LSR}$ $\sim$$+$30 km s$^{-1}$ for Object 2). 
In addition, the CO and SiO emission in both objects show large line FWHMs of 8--14 km s$^{-1}$ (thermal velocity is $\sim$0.1 km s$^{-1}$ for 20 K CO gas). 
Their FWHMs are large than those of observed in quiescent dark clouds \citep[$\lesssim$1 km s$^{-1}$, e.g., ][]{Lar81}. 

No other emission line was detected although the present spectral setup covers transitions of dense gas tracers such as H$^{12}$CO$^+$(4-3, $E_u$= 43 K), H$^{13}$CN(4-3, $E_u$= 41 K), and SO($N_J$ = 8$_{8}$--7$_{7}$, $E_u$= 88 K). 
Several strong transitions of \ce{CH3OH} and \ce{SO2} with $E_u$ ranging from 40 K to 500 K are also covered but not detected. 
A typical rms noise level at the frequencies of these emission lines is similar to those of the CO and SiO lines ($\sim$0.05). 
No continuum emission was detected in both objects despite the presence of deep ice and dust absorption bands. 
A 1$\sigma$ upper limit on the beam-averaged flux is about 0.1 mJy/beam. 

Because the target sources are located in the direction of the Crux-Scutum arm (Fig.~\ref{direction}), multiple different velocity components of the CO gas are detected within the field-of-view (Figures \ref{obj1}--\ref{obj2} b, e). 
These extended emission components do not spatially overlap with the positions of Object 1 and 2. 
In addition, their systemic velocities are different from those of the surrounding CO gas. 
Full channel maps in the frequency region of the CO(3--2) emission are shown in Appendix \ref{sec_app_chanmap}. 
It is clearly seen from the channel maps that the CO emission associated with Object 1 and 2 are spatially and kinematically separated from the surrounding CO gas. 
Kinematic properties of CO gas in the observed regions are further discussed in Section \ref{sec_disc3}.

\section{Discussion} \label{sec_disc} 
\subsection{Physical and chemical characteristics of the icy objects} \label{sec_disc1} 
The infrared characteristics of two peculiar icy objects are summarized as 
(i) deep and rich ice (H$_2$O, CO$_2$, CO, CH$_3$OH, and XCN) and dust (silicate) absorption features, 
(ii) strong and localized reddening at the position of the infrared point-source, 
(iii) SED peak at 5 $\mu$m and no excess at longer wavelengths, 
as reported in \citet{Ona21}

Their submillimeter characteristics revealed by the present observations are  
(iv) compact SiO and CO emission at the position of the infrared point-source, 
(v) broad line widths of SiO and CO emission, 
(vi) systemic velocities clearly separated from the surrounding CO clouds, 
(vii) no continuum emission despite the presence of deep ice/dust absorption. 
We here discuss a possible nature of Object 1 and 2 based on their submillimeter and infrared characteristics.

\subsection{SiO/CO ratio} \label{sec_disc2} 
SiO is often used as a tracer of shocked gas. 
It is believed that energetic processes in shock regions such as sputtering or photolysis release silicon or silicon-bearing molecules from dust grains, which leads to the formation of SiO \citep[e.g.,][]{Sch97_SiO, Tab20}. 
We here compare the column density ratios of SiO and CO ($N_{\mathrm{SiO}}$/$N_{\mathrm{CO}}$) in Object 1 and 2 with those measured in various astronomical environments (Table \ref{tab_SiO_CO}). 

The $N_{\mathrm{SiO}}$ and $N_{\mathrm{CO}}$ of the two objects are estimated based on the standard treatment of optically-thin emission under the local thermodynamic equilibrium (LTE). 
The details of the calculation are described in Section \ref{sec_app_N}. 
The rotation temperatures of SiO and CO gas are unknown. 
We here assume rotation temperatures of 20--50 K and estimate the column densities and the uncertainties associated with the temperature assumption. 
Since no strong high-excitation lines of high-temperature gas tracers such as CH$_3$OH, SO, SO$_2$ are detected, we exclude the possibility of $T_\mathrm{rot} \gtrsim80$\,K. 

The $N_{\mathrm{CO}}$ is derived to be (2.2--1.9) $\times$ 10$^{15}$ cm$^{-2}$ with $T_\mathrm{rot}$ = 20--50 K for Object 1 and (1.5--1.3) $\times$ 10$^{15}$ cm$^{-2}$ for Object 2. 
The $N_{\mathrm{SiO}}$ is derived to be (9.4--2.5) $\times$ 10$^{12}$ cm$^{-2}$ for Object 1 and (4.8--1.2) $\times$ 10$^{12}$ cm$^{-2}$ for Object 2, respectively. 
Thus, the $N_{\mathrm{SiO}}$/$N_{\mathrm{CO}}$ ratio is (2--4) $\times$ 10$^{-3}$ in Object 1 and (0.9--3) $\times$ 10$^{-3}$ in Object 2, respectively. 
If we assume an even lower $T_\mathrm{rot}$ of 10 K, then we obtain the $N_{\mathrm{SiO}}$/$N_{\mathrm{CO}}$ ratio of several $\times$ 10$^{-2}$ in both objects. 

This high SiO/CO ratio is comparable with those observed in shocked regions associated with protostellar outflows/jets, which are typically 10$^{-4}$--10$^{-2}$ \citep[e.g.,][]{Bac91, Taf10, Dut24}. 
The present SiO/CO ratios are significantly higher than those observed in dark clouds and protostellar envelopes, which are about $\sim$10$^{-8}$ and $\sim$10$^{-6}$, respectively \citep[e.g.,][]{Ziu89, Ger14}. 
Large line widths of CO and SiO emission in both objects ($\sim$10 km s$^{-1}$) support their highly non-thermal and turbulent nature. 
High SiO/CO ratios as well as large line widths observed in Object 1 and 2 may suggest that they are associated with shocked gas. 

Alternatively, the high SiO/CO ratio may be related to the circumstellar chemistry in evolved stars, since SiO and CO are one of the most abundant detectable molecules in the envelope of oxygen-rich AGB stars \citep[e.g.,][]{Dur99}. 
They exhibit intense SiO and CO lines even for thermal emission, and their SiO/CO ratios typically range from 10$^{-3}$ to 10$^{-1}$ as in Table \ref{tab_SiO_CO} \citep[e.g.,][]{Buj94, Gon03}. 
In addition, SiO and CO lines in oxygen-rich AGB stars generally show a large line width \citep[$>$10 km s$^{-1}$, e.g.,][]{Jus96, vanSan18, Olo22}. 
Although some AGB stars are known to exhibit double-peak or non-Gaussian spectral line profiles \citep[e.g.,][]{Kem03, DeB10}, such features are difficult to identify in Object 1 and 2 due to the limited spectral resolution and sensitivity of the present data. 
The possibility of an evolved star as a nature of the present icy objects will be further discussed in Section \ref{sec_disc5b} in conjunction with their infrared characteristics.

\begin{deluxetable}{ l c c}
\tablecaption{$N_{\mathrm{SiO}}$/$N_{\mathrm{CO}}$ in various sources   \label{tab_SiO_CO} }
\tabletypesize{\small} 
\tablehead{
\colhead{Source} & \colhead{$N_{\mathrm{SiO}}$/$N_{\mathrm{CO}}$}  & \colhead{Reference}   \\
}
\startdata 
Object 1 $\&$ 2          & $\sim$1 $\times$ 10$^{-3}$        & this work   \\
Shock regions           & $\sim$10$^{-4}$--10$^{-2}$        & 1, 2, 3   \\
Protostellar envelopes  & $\sim$10$^{-6}$                   & 4   \\
Dark clouds             & $\sim$10$^{-8}$                   & 5   \\
AGB stars               & $\sim$10$^{-3}$--10$^{-1}$        & 6, 7   \\
\enddata
\tablecomments{
$^{1}$\citet{Bac91}, 
$^{2}$\citet{Taf10}, 
$^{3}$\citet{Dut24}, 
$^{4}$\citet{Ger14}, 
$^{5}$\citet{Ziu89},
$^{6}$\citet{Buj94},
$^{7}$\citet{Gon03}
}
\end{deluxetable}

\subsection{Systemic velocities and kinematic distances} \label{sec_disc3} 
Systemic velocities of point-like CO and SiO emission associated with the icy objects are clearly different from those of surrounding gas as shown in Figures \ref{obj1} and \ref{obj2}. 
We estimate kinematic distances to the icy objects and the surrounding gas using the kinematic distance calculator\footnote{\url{http://bessel.vlbi-astrometry.org/revised_kd_2014}}, which is based on the method described in \citet{Rei09} and the Galactic parameters from \citet{Rei14}. 

For Object 1, the surrounding gas has systemic velocities between $-$70 and $-$45 km s$^{-1}$, which correspond to the kinematic distances of 3.4--5.5 kpc (near) and 5.5--7.9 kpc (far). 
Molecular gas located in the Crux-Scutum arm region would account for such ranges of kinematic distances as seen in Fig.~\ref{direction}. 
On the other hand, the systemic velocity of Object 1 ($-$26 km\,s$^{-1}$ in CO) corresponds to the kinematic distance of 2.0 $\pm$ 0.5 kpc (near) and 9.3 $\pm$ 0.5 kpc (far). 
If Object 1 is co-moving with the Galactic rotation and the kinematic distance is correct, the object would be located either in the Crux-Scutum arm or in the near side of the Carina-Sagittarius arm. 

Similarly, for Object 2, the bright components of the surrounding gas have systemic velocities between $-$65 and $-$20 km s$^{-1}$, which correspond to the kinematic distances of 1.9--5.5 kpc (near) and 5.5--9.4 kpc (far). 
The other bright components with positive systemic velocities (2.5 and 6.0 km $s^{-1}$) have the kinematic distance of 11.3--11.6 $\pm$ 0.5 kpc (no near/far ambiguity). 
These kinematic distances would correspond to the molecular gas located in the Crux-Scutum arm region (or partly in the near side of the Carina-Sagittarius arm). 
On the other hand, the systemic velocity of Object 2 ($+$29 km\,s$^{-1}$ in CO) corresponds to the kinematic distance of 13.4 $\pm$ 0.6 kpc (no near/far ambiguity). 
Such a kinematic distance corresponds to the inter-arm region between the Crux-Scutum and the Carina-Sagittarius arms as seen in Figure \ref{direction}.

A clear velocity and spatial separation of the molecular gas associated with Object 1 and 2 from those of the surrounding line-of-sight CO gas would suggest that they are isolated and not associated with extended CO gas. 
Furthermore, although Object 1 and 2 show similar infrared and submillimeter characteristics and they are located in the similar direction in the sky (3.3 arcminutes separation), their systemic velocities are different from each other, suggesting that they are kinematically unrelated.

\subsection{Luminosities} \label{sec_disc4} 
\begin{figure}[tp!]
\begin{center}
\includegraphics[width=8.7cm]{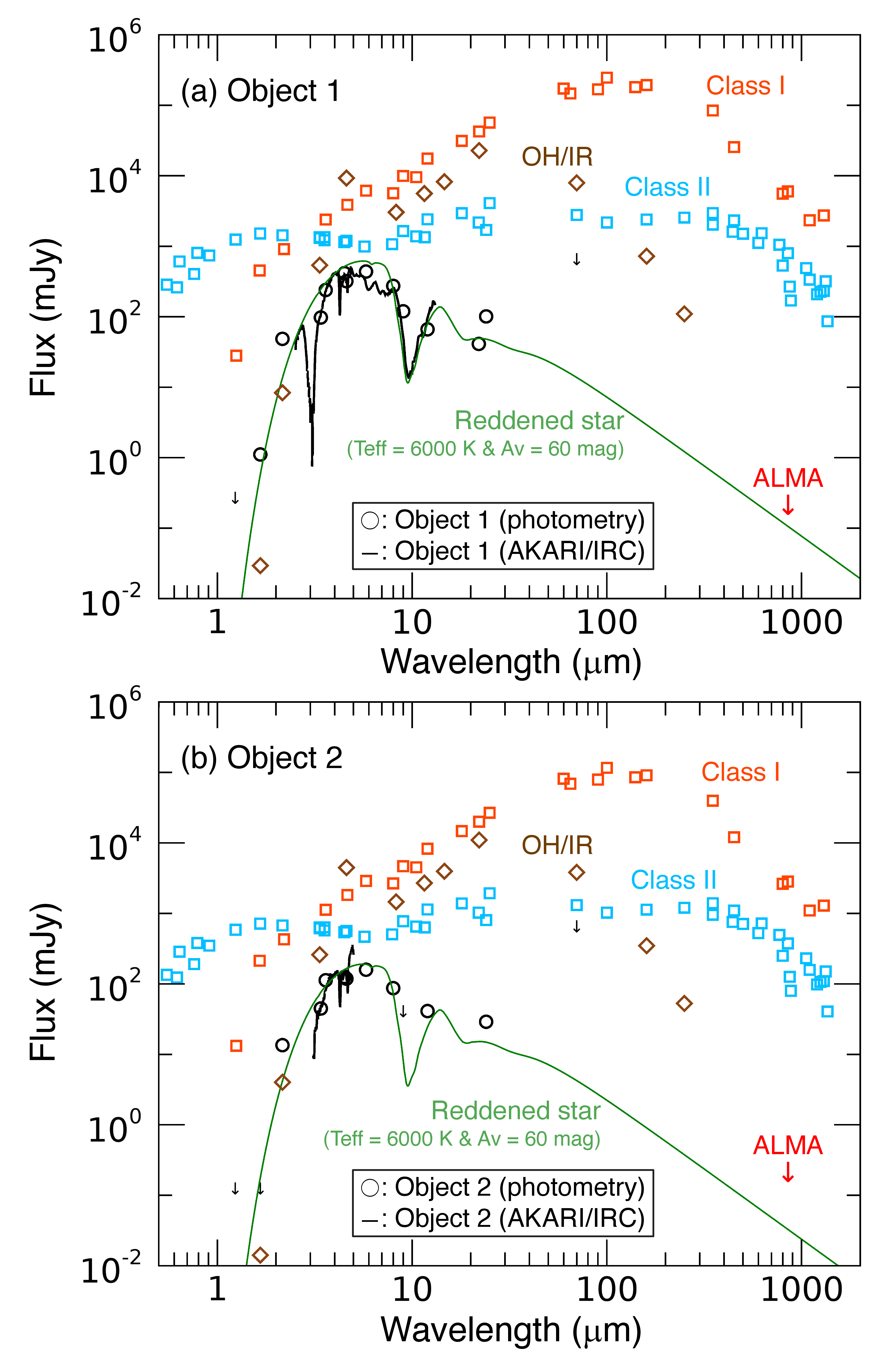}
\caption{
SEDs for Object 1 (a) and Object 2 (b), where the black circles represent the photometric data from the 2MASS, WISE, Spitzer, AKARI, and Herschel \citep[see][and references therein]{Ona21}. 
The black downward arrows indicate photometric upper limits. 
The black solid line represents the AKARI spectrum. 
The red downward arrows indicate the upper limit obtained with the present ALMA observations. 
Typical SEDs of Class I and II YSOs in Taurus are plotted by orange and blue squares \citep[][]{Ell19, Rib17}. 
The brown diamonds represent the photometric data of an OH/IR star, OH 26.5+0.6, obtained by 2MASS, MSX, WISE, and Herschel, whose fluxes are scaled to fit the distance to Object 1 or 2 \citep{Skr06, Ega03, Wri10, Mol16}. 
The green lines represent the synthesized spectrum of a reddened star with the effective temperature (T$\mathrm{_{eff}}$) of 6000 K and the visual extinction of 60 mag. 
The flux of Class I/II YSOs and reddened star are arbitrary scaled for better visibility. 
See Section \ref{sec_disc5} for details. 
}
\label{SED}
\end{center}
\end{figure}

We estimate the bolometric luminosities of Object 1 and 2 based on the above kinematic distances and available photometric data. 
We have integrated their photometric data from 2 to 20 $\mu$m, where the most energy is emitted. 
Their SEDs are shown in Figure \ref{SED}. 
For Object 1, the estimated luminosity is 30 $L_\odot$ for the near distance (2.0 kpc) and 750 $L_\odot$ for the far distance (9.3 kpc), while for Object 2, it is 500 $L_\odot$ for the distance of 13.4 kpc. 

In the following discussion, we assume the far distance for Object 1, because it shares a common submillimeter and infrared characteristics with Object 2. 
The non-thermal motion of molecular gas as well as the high SiO/CO ratio commonly observed in Object 1 and 2 would suggest that the present ALMA observations probe a similar physical scale for the objects with comparable luminosities. 
Thus, it is more likely that Object 1 is located at the distant region as in the case of Object 2. 
Their possible locations are shown in Figure \ref{direction}. 
At the assumed distance of 9--13 kpc, their bolometric luminosities correspond to those of intermediate-mass stars. 


\subsection{Gas column densities and constraint on the source size} \label{sec_disc5} 
A large dust extinction is suggested from deep dust/ice absorption of Object 1 and 2, however, no continuum emission is detected in both sources. 
Since the infrared absorption spectroscopy probes the line of sight in a pencil-like beam, the non-detection of continuum emission at submillimeter wavelengths suggest a large beam dilution. 

Table \ref{tab_N_H2} summarizes the \ce{H2} column densities ($N_{\mathrm{H_2}}$) estimated by various methods. 
First, we estimate $N_{\mathrm{H_2}}$ by using the peak optical depth of the 9.7 $\mu$m silicate dust absorption ($\tau_{\mathrm{9.7}}$) and the 3.0 $\mu$m H$_2$O ice absorption ($\tau_{\mathrm{3.0}}$), which are measured to be 2.9 $\pm$ 0.3 and  4.5 $\pm$ 0.5 for Object 1 \citep{Ona21}. 
Those for Object 2 are unknown because the bottom of the absorption is below the noise level. 
Since ice absorption depths of Object 2 is comparable with that of Object 1, we here assume the same absorption depth for both objects. 
Details of the conversion from $\tau_{\mathrm{9.7}}$ and $\tau_{\mathrm{3.0}}$ to $N_{\mathrm{H_2}}$ are described in Section \ref{sec_app_tau97} in Appendix. 

The above $N_{\mathrm{H_2}}$ is compared with those derived from the submillimeter data. 
First, the CO column density derived in Section \ref{sec_disc2} is converted to $N_{\mathrm{H_2}}$ assuming a standard CO/\ce{H2} ratio of 10$^{-4}$ in a solar metallicity \citep[e.g.,][]{Hir17,Bis21}. 
Next, we estimate an upper limit on the $N_{\mathrm{H_2}}$ based on the non-detection of continuum emission. 
Details of these calculation are explained in Section \ref{sec_app_h2} in Appendix. 

The $N_{\mathrm{H_2}}$ values derived by the ALMA CO data are significantly lower than that from the dust/ice absorption band as tabulated in Table \ref{tab_N_H2}. 
This discrepancy may suggest a large beam dilution of the ALMA data as mentioned above. 
To obtain the $N_{\mathrm{H_2}}$ that is consistent between the dust/ice absorption and CO emission, we need to assume the beam dilution factor of 10000 (Object 1) and 14000 (Object 2), which means that the source is smaller than the beam size by a factor of 100 or 120.  
At the assumed distances of 9.3 kpc and 13.4 kpc for Object 1 and 2, the present ALMA beam size (0$\farcs$8) corresponds to 7400 au and 10700 au. 
Considering the above beam dilution, the expected physical sizes of Object 1 and 2 are 74 au and 90 au, respectively. 

Similarly, an upper limit on $N_{\mathrm{H_2}}$ from the non-detection of the ALMA continuum is lower than the $N_{\mathrm{H_2}}$ from the dust/ice absorption by a factor of 100. 
We thus need to assume at least a 10 times smaller source size for the dust emitting region compared to the beam size, which corresponds to 740 au and 1070 au for Object 1 and 2. 
These continuum-based upper limits on the source size are consistent with the aforementioned size estimate with the CO data. 
We note that the column density estimate with CO assumes optically-thin emission, which is still uncertain. 
However, with the continuum-based upper limits, we can still constrain the source size to be less than $\sim$1000 au. 

The point-like and faint emission of CO, as well as the non-detection of the dust emission, would suggest that the physical size of Object 1 and 2 is around 100-1000 au. 
Within the known interstellar objects with ice absorption bands, such a physical size corresponds to those of protoplanetary disks \citep[e.g.,][]{Wil11}. 


\begin{deluxetable}{ l c c}
\tablecaption{Estimated $N_{\mathrm{H_2}}$ and source size  \label{tab_N_H2} }
\tabletypesize{\footnotesize} 
\tablehead{
\colhead{Method} & \colhead{$N_{\mathrm{H_2}}$ (cm$^{-2}$)}  & \colhead{Size}   
}
\startdata 
Object 1  & & \\
Dust/ice absorption                                             & 2 $\times$ 10$^{23}$        & \nodata                    \\
Dust emission ($T_{d}$ = 20--50 K)\tablenotemark{a}             & $<$5--2 $\times$ 10$^{21}$  & \nodata                     \\
Dust emission ($\times$ 100 beam dilution) \tablenotemark{a}    & $<$5--2 $\times$ 10$^{23}$  & $<$740 au                 \\
CO emission ($T_{\rm rot}$ = 20--50 K $\&$ LTE)                 & 2.2--1.9 $\times$ 10$^{19}$ & $<$7400 au                  \\
CO emission ($\times$ 10000 beam dilution)                      & $\sim$2 $\times$ 10$^{23}$  & 74 au                       \\
\hline
Object 2  & & \\
Dust/ice absorption                                             & 2 $\times$ 10$^{23}$        & \nodata                        \\
Dust emission ($T_{d}$ = 20--50 K)                              & $<$5--2 $\times$ 10$^{21}$  & \nodata                      \\
Dust emission ($\times$ 100 beam dilution)                      & $<$5--2 $\times$ 10$^{23}$  & $<$1070 au                  \\
CO emission ($T_{\rm rot}$ = 20--50 K $\&$ LTE)                 & 1.5--1.3 $\times$ 10$^{19}$ & $<$10700 au              \\
CO emission ($\times$ 14000 beam dilution)                      & $\sim$2 $\times$ 10$^{23}$  & 90 au                     \\
\enddata
\tablenotetext{a}{A 3-$\sigma$ upper limit from the continuum data. }
\end{deluxetable}

\subsection{Comparison with the characteristics of known icy objects} \label{sec_disc5} 
Deep ice absorption features are observed in embedded Class I YSOs, Class II YSOs with edge-on disks, field stars behind dense clouds, mass-losing oxygen-rich evolved stars, and extragalaxies (see Section \ref{sec_intro} for references). 
Object 1 and 2 would not be an extragalactic source according to their systemic velocities. 

\subsubsection{Prestellar/Protostellar objects} \label{sec_disc5a} 
Figure \ref{SED} shows SEDs of Object 1 and 2 together with the typical SEDs of Class I and Class II YSOs. 
Also plotted is the synthesized spectrum of a reddened star with an arbitrary scaled flux. 
The spectrum was calculated by applying the interstellar extinction curve of \citet{Dra03} with $A_V$ = 60 mag and $R_V$ = 4.0 to a 6000 K blackbody curve \footnote{Extinction curve was calculated by using the dust extinction package \citep{Gor24}.}. 
The SED shape of Object 1 and 2 are closer to that of a reddened star, rather than those of Class I/II YSOs as shown in the figure. 
However, if the estimated kinematic distance to the objects are correct, the continuum source (i.e., background star) need to be unrealistically luminous, because of the large extinction ($A_V$ $\sim$100 mag, $A_{K_\mathrm{S}}$ $\sim$11 mag) and distance ($>$9-13 kpc, distance modulus $>$15 ) to the objects; cf., the $K_\mathrm{S}$-band magnitudes of Object 1 and 2 are 10.4 and 11.7 mag, respectively \citep{Ona21}. 

Class I YSOs are often associated with protostellar outflows and jets that trigger shock chemistry, and also show deep ice absorption, but such objects are much brighter in mid-/far-infrared and submillimeter wavelengths and their SEDs deviate from those of Object 1 and 2 as shown in Figure \ref{SED}. 
The inferred small source size ($\sim$100 au) of the objects hints at the possibility of Class II YSOs with edge-on disks. 
However, even for Class II YSOs, most energy is emitted in the longer wavelength side \citep[e.g.,][]{Cra08}, and their SEDs do not agree with those of the present objects (Fig.~\ref{SED}). 
If the absorbing disk is very clumpy, and seen together with stellar or disk inner radiation, then the mid-/far-infrared and submillimeter emission could be relatively reduced and may be reconciled with the observations. 
If this is the case, similar objects could have eluded past photometric surveys of YSOs. 

We note that the two objects in this work were discovered by the slitless spectroscopic survey toward 22 fields with various galactic longitudes in the Galactic plane, where a 10$\arcmin$ $\times$ 10$\arcmin$ region was observed for each field \citep{Ona21}. 
The present two objects were found in the direction of the Crux arm, but no similar object was identified in the remaining 21 fields. 
The peculiarity in their observed direction, along with the rarity of their existence within the scope of the conducted slitless spectroscopic survey, suggests that caution should be exercised when considering how many similar objects exist. 

\subsubsection{Evolved stars}  \label{sec_disc5b} 
High-mass-loss-rate oxygen-rich AGB stars, known as OH/IR stars, are another potential site of ice formation. 
Owing to the presence of a thick circumstellar envelope, some OH/IR stars exhibit the 3 $\mu$m H$_2$O ice absorption feature in addition to the deep 10 $\mu$m dust absorption \citep[e.g.,][]{Syl99, Jus06, Suh13}. 

As discussed in Section \ref{sec_disc2}, the SiO/CO ratio in Object 1 and 2 are not very different from those observed in oxygen-rich AGB stars. 
In addition, the low CO line intensities in the present icy objects do not contradict the AGB case. 
The ALMA spectra in Figure \ref{obj1}--\ref{obj2} (f) probe nearly 10000 au region around the target sources if their kinematic distances are correct. 
Submillimeter single-dish observations towards nearby ($d$ = 0.6--2.3 kpc) OH/IR stars have reported that the peak brightness temperature of the CO(3-2) line is about 0.2--2.5 K for a $\sim$10000--30000 au region, which is comparable with the intensities of the present icy objects \citep{Jus96, DeB10}. 
Regarding the continuum strength, \citet{Olo22} have reported the ALMA Band 7 observations towards $\sim$20 OH/IR stars in the Galactic center region ($d$ = 8.2 kpc). 
For some bright sources, the reported 339 GHz continuum flux is about 15 mJy, which will be detectable with the present observations even at $\sim$10--20 kpc. 
However, the 339 GHz continuum was not detected in many other sources and their fluxes are below $\sim$2 mJy. 
Such submillimeter-faint OH/IR stars may not be detectable with the present observations. 
Therefore, we cannot fully rule out the possibility of AGB stars as a nature of the present icy objects solely from their submillimeter properties. 

On the other hand, the infrared properties of Object 1 and 2 are different from those of OH/IR stars. 
Figure \ref{SED} shows the SED of a well-studied OH/IR star, OH 26.5+0.6. 
The photometric data is taken from the NASA/IPAC Infrared Science Archive and based on 2MASS, MSX, WISE All-sky, and Herschel Hi-GAL data \citep{Skr06, Ega03, Wri10, Mol16}. 
OH 26.5+0.6 is located at the distance of 1.3 kpc \citep{vanLan90}. 
The plotted SED of the source is scaled to fit the distance to Object 1 or 2. 
As shown in the figure, the lack of mid- to far-infrared excess in the present icy objects does not match the characteristic SED of the OH/IR star. 
Due to the lack of the long-wavelength excess, the bolometric luminosities of Object 1 and 2 are roughly an order of magnitude smaller than those of typical OH/IR stars \citep[e.g.,][]{Lep95,Olo22}. 
Furthermore, the ice species detected in OH/IR stars is only H$_2$O \citep[e.g.,][]{Syl99}. 
Other ice species, such as CO, CO$_2$, and CH$_3$OH as detected in Object 1 and 2, are not reported in OH/IR stars. 
Perhaps, this would be due to the relatively warm temperature of the circumstellar envelope of OH/IR stars compared to dense molecular clouds, which inhibit the adsorption of highly-volatile species. 
The time variation of flux is another important characteristic of evolved stars, but such a variability is not detected in the present objects. 

Other types of AGB stars also do not match the infrared characteristics of Object 1 and 2. 
Carbon-rich AGB stars (carbon stars) are another class of high-mass-loss-rate evolved stars. 
They show absorption bands due to gaseous HCN (3.1 $\mu$m) and C$_2$H$_2$ (3.1 and 3.8 $\mu$m), while they do not show the 10 $\mu$m silicate dust absorption \citep[e.g.,][]{Yam00, Mat02, Gro07}. 
Normal M-type AGB stars and S-type AGB stars generally show the 10 $\mu$m silicate dust feature in emission due to their thinner envelope \citep[e.g.,][]{Syl99, Ona02, Her05, Hony09}. 
None of the above infrared features are observed in the present objects. 

\section{Summary} \label{sec_sum} 
We report the results of ALMA observations towards two peculiar objects with unknown nature, which were serendipitously discovered by the previous infrared spectroscopic survey of the Galactic plane with the AKARI satellite. 
Both objects show strong dust and ice absorption features. 
We detect very compact emission of CO(3-2) and SiO(8-7) at the positions of the icy objects. 
Because they are located in the direction of the Crux-Scutum arm, multiple different velocity components of the CO gas are detected around them. 
However, the systemic velocities and spatial distribution of molecular gas associated with the icy objects are clearly separated from the surrounding CO clouds, suggesting that they are isolated from the line-of-sight clouds. 
Although they show similar infrared and submillimeter characteristics and they are located in similar directions in the sky (3.3 arcminutes separation), their systemic velocities are different from each other, suggesting that they are kinematically unrelated. 
Their kinematic distances are estimated to be 9.3 kpc and 13.4 kpc, respectively. 
With these kinematic distances, the bolometric luminosities of the icy objects are estimated to be 750 $L_\odot$ and 500 $L_\odot$. 

The observed large column density ratios of SiO/CO gas ($\sim$10$^{-3}$) in both objects are comparable with those observed in shocked regions associated with protostellar outflows and jets. 
Along with the broad line widths (8-14 km s$^{-1}$) of their CO and SiO emission, it is likely that they are associated with shocked gas. 
Although a large dust extinction ($A_V$ $\sim$100 mag) is expected from their deep dust/ice absorption features, no dust continuum emission is detected, which would suggest a large beam dilution effect due to the compact size of the sources. 
Based on the comparison of the gas column density estimates with infrared dust/ice absorption, submillimeter CO gas emission, and submillimeter dust emission, we estimate the source size of the icy objects to be $\sim$100-1000 au. 
Despite their deeply embedded characteristics, their SEDs show a peak around 5 $\mu$m and decrease toward longer wavelengths. 

Deeply dust-enshrouded oxygen-rich AGB stars (OH/IR stars) are another potential site of ice formation. 
We cannot fully rule out the possibility of OH/IR stars solely from the submillimeter properties of the present icy objects, however, their infrared characteristics are very different from those of known OH/IR stars with ice absorption. 

These characteristics, (i) rich ice absorption features, (ii) large visual extinction, (iii) lack of mid-infrared and submillimeter excess emission, (iv) very compact source size, (v) SiO-dominated broad molecular line emission, and (vi) isolation, cannot easily be accounted for by any of known interstellar icy sources. 
They may represent a previously unknown or rare type of isolated icy objects. 
Future high-spatial-resolution and high-sensitivity observations as well as detailed SED modeling is required. 
An upcoming near-infrared spectroscopic survey with SPHEREx \citep{Ash23} may detect more similar sources.

\section*{acknowledgments}
This paper makes use of the following ALMA data: ADS/JAO.ALMA$\#$2021.1.01126.S. 
ALMA is a partnership of ESO (representing its member states), NSF (USA) and NINS (Japan), together with NRC (Canada), MOST and ASIAA (Taiwan), and KASI (Republic of Korea), in cooperation with the Republic of Chile. 
The Joint ALMA Observatory is operated by ESO, AUI/NRAO and NAOJ. 
This work has made use of the Cologne Database for Molecular Spectroscopy. 
This work was supported by JSPS KAKENHI grant Nos. JP20H05845, JP21H01145, and JP24K07087. 
T. S. was supported by Leading Initiative for Excellent Young Researchers, MEXT, Japan.
Finally, we would like to thank an anonymous referee for insightful comments, which substantially improved this paper. 



\software{CASA \citep{McM07})}




\appendix
\section{Derivation of molecular column densities} \label{sec_app_N} 
Column densities of CO and SiO are estimated as follows based on the standard treatment of optically thin emission in the LTE \citep[e.g.,][]{Yam17}. 

\begin{equation}
N = \frac{3 k Q(T_{\mathrm{rot}}) \int T_{\mathrm{b}} dV}{8 \pi^{3} \nu S \mu^{2}} \Bigg{\{} 1 - \frac{\exp(h \nu / kT_{\mathrm{rot}})-1}{\exp(h \nu / kT_{\mathrm{bg}})-1} \Bigg{\}}^{-1} \exp\Bigg{(}\frac{E_u}{kT_{\mathrm{rot}}}\Bigg{)}  \label{Eq1}, 
\end{equation}
where 
$N$ is the total column density, 
$k$ is the Boltzmann constant, 
$T_{\mathrm{rot}}$ is the rotational temperature, 
$T_{\mathrm{bg}}$ is the cosmic microwave background temperature (2.73 K) , 
$Q(T_{\mathrm{rot}})$ is the partition function at $T_{\mathrm{rot}}$, 
$\int T_{\mathrm{b}} dV$ is the integrated intensity estimated from the observations, 
$\nu$ is the transition frequency, 
$S$ is the line strength, $\mu$ is the dipole moment, 
$h$ is the Planck constant, 
and 
$E_{u}$ is the upper state energy. 
All the spectroscopic parameters required in the analysis are extracted from the CDMS database.

\section{Derivation of the H$_2$ column density from $\tau_{\mathrm{9.7}}$ and $\tau_{\mathrm{3.0}}$} \label{sec_app_tau97} 
The relation between $A_V$ and $\tau_{\mathrm{9.7}}$ reported for Galactic dense cores is 
\begin{equation}
A_V = \frac{\tau_{\mathrm{9.7}} - (0.12 \pm 0.05)}{0.21 \pm 0.02} \times 8.8, \label{Eq_Avtau97} 
\end{equation}
according to \citet{Boo11} (assuming $A_V$/$A_K$ = 8.8). 
Applying this relation to Object 1 ($\tau_{\mathrm{9.7}}$ = 2.9 $\pm$ 0.3), we obtain $A_V$ = 117$^{+28}_{-24}$ mag. 
%

Similarly, the relation between $A_V$ and $\tau_{\mathrm{3.0}}$ in \citet{Boo11} is
\begin{equation}
A_V = \frac{\tau_{\mathrm{3.0}} + (0.15 \pm 0.13)}{0.45 \pm 0.05} \times 8.8. \label{Eq_Avtau30} 
\end{equation}
Applying this relation to Object 1 ($\tau_{\mathrm{3.0}}$ = 4.5 $\pm$ 0.5), we obtain $A_V$ = 91$^{+25}_{-20}$ mag. 

By averaging the above two estimates, we use $A_V$ = 104 mag in the subsequent calculation. 
We assume that the object is embedded in a dense core. 
In this case, the infrared absorption spectroscopy probes only the foreground component relative to the continuum source. 
Therefore, the above $A_V$ value is doubled to compare with submillimeter data, which probe the total column density in the line of sight. 

To estimate $N_{\mathrm{H_2}}$ from $A_V$, we use a $N_{\mathrm{H}}$/($A_{B}$ $-$ $A_{V}$) conversion factor of 5.8 $\times$ 10$^{21}$ cm$^{-2}$ mag$^{-1}$ \citep{Boh78}. 
We use a slightly high $A_{V}$/($A_{B}$ $-$ $A_{V}$) ratio of 4 for dense clouds \citep{Whi01b}, and assume that all the hydrogen atoms are in the form of H$_2$. 
Then we obtain $N_{\mathrm{H_2}}$/$A_{V}$ = 7.3 $\times$ 10$^{20}$ cm$^{-2}$ mag$^{-1}$, which was used for the estimate of $N_{\mathrm{H_2}}$ tabulated in Table \ref{tab_N_H2}.

\section{Derivation of the H$_2$ column density from dust continuum} \label{sec_app_h2} 
Based on the standard treatment of optically thin dust emission, we apply the following equation to calculate an upper limit of the H$_2$ column density ($N({\ce{H2}})$):
\begin{equation}
N({\ce{H2}}) = \frac{F_{\nu} / \Omega}{2 \kappa_{\nu} B_{\nu}(T_{d}) Z \overline{M}_w m_{\mathrm{H}}} \label{Eq2}, 
\end{equation}
where $F_{\nu}/\Omega$ is the continuum flux density per beam solid angle as estimated from the observations, $\kappa_{\nu}$ is the mass absorption coefficient of dust grains coated by thick ice mantles taken from \citet{Oss94} and we use 2.14 cm$^2$ g$^{-1}$ at 853 $\mu$m, $T_{d}$ is the dust temperature, $B_{\nu}(T_{d})$ is the Planck function, $Z$ is the dust-to-gas mass ratio and we use a canonical value of 0.008 for the solar neighborhood, $\overline{M}_w$ is the mean atomic mass per hydrogen \citep[1.41,][]{Cox00}, and $m_{\mathrm{H}}$ is the hydrogen mass.

\section{Channel maps of CO(3-2)} \label{sec_app_chanmap} 
\begin{figure*}[tp!]
\begin{center}
\includegraphics[width=16.0cm]{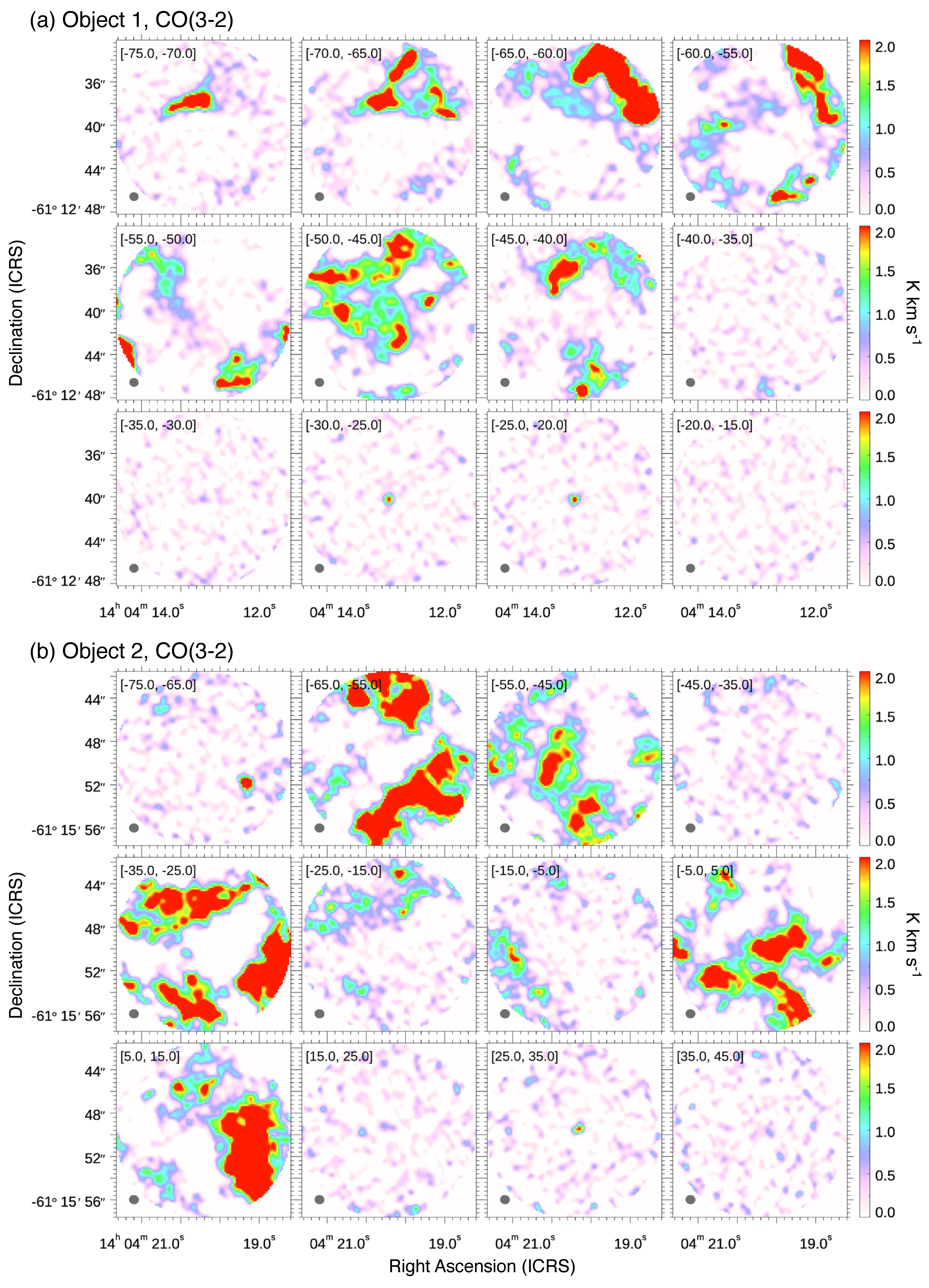}
\caption{
ALMA CO(3--2) channel maps for the region around Object 1 (a) and 2 (b). 
The plotted velocity ranges are from $-$75 to $-$15 km s$^{-1}$ with an interval of 5 km s$^{-1}$ for Object 1, and from $-$75 to 45 km s$^{-1}$ with an interval of 10 km s$^{-1}$ for Object 2. 
The integrated velocity ranges are shown at the upper left corner of each panel. 
Object 1 and 2 are located at the image center as a point source, and they are visible in the panels of [$-$30.0, $-$25.0] and [$-$25.0, $-$20.0] for (a) and [25.0, 35.0] for (b), respectively. 
The gray ellipse represents the synthesized beam size. 
}
\label{chanmap}
\end{center}
\end{figure*}

Figure \ref{chanmap} shows the CO(3--2) channel maps of the region around Object 1 and 2. 
These are constructed by integrating the data cube in the velocity ranges from $-$75 to $-$15 km s$^{-1}$ with an interval of 5 km s$^{-1}$ for Object 1, and from $-$75 to 45 km s$^{-1}$ with an interval of 10 km s$^{-1}$ for Object 2, respectively. 

\end{document}